\documentstyle[aaspp4]{article}

\input{psfig}

\begin{document}
 
\begin{center}\uppercase{{\sc Self Similar Spherical Collapse Revisited:
a Comparison between Gas and Dark Matter Dynamics}}
\end{center}

\author{\sc Romain    Teyssier, Jean-Pierre   Chièze}
\affil{CEA, DSM/DAPNIA/Service d'Astrophysique, CE-Saclay, 
  F-91191 Gif-sur-Yvette, Cedex, France}

\author{\sc Jean-Michel Alimi}
\affil{Laboratoire d'Astrophysique Extragalactique et de Cosmologie,
CNRS URA 173, Observatoire de Paris-Meudon, 92195 Meudon, France}

\begin{abstract}
  We  reconsider the collapse of  cosmic  structures in an Einstein-de
  Sitter Universe,   using the  self   similar initial  conditions  of
  Fillmore  \& Goldreich (1984).  We  first derive a new approximation
  to describe the dark matter dynamics in  spherical geometry, that we
  refer to the ``fluid approach''.  This method  enables us to recover
  the  self--similarity  solutions of Fillmore   \& Goldreich for dark
  matter.  We derive also new self--similarity  solutions for the gas.
  We thus compare  directly gas and dark  matter dynamics, focusing on
  the differences due  to their different dimensionalities in velocity
  space.  This work may have interesting consequences for gas and dark
  matter distributions in large  galaxy clusters, allowing to  explain
  why the  total  mass profile is always  steeper  than the  X-ray gas
  profile.  We  discuss  also the shape  of  the  dark  matter density
  profile    found in N--body  simulations in    terms of  a change of
  dimensionality in    the dark matter  velocity space.     The stable
  clustering  hypothesis has been finally   considered in the light of
  this analytical approach.
\end{abstract}


\section{INTRODUCTION}

The  study of  cosmic  structures formation  within the  gravitational
instability picture  has  proven to be   a usefull tool  over the last
decade to connect the initial power spectrum with the density profiles
in dark matter halos.  Gunn  \& Gott (1972) derived analytical density
profiles obtained  during the collapse  of  a self-similar dark matter
density  perturbation.  This  so--called  secondary infall  theory was
further   extended  and refined by   Fillmore \&  Goldreich (1984) and
Bertschinger (1985).  The latter also included a detailed treatment of
gas collapse, and  showed that gas  and dark matter  distributions are
similar in the special case that he studied.  Hoffman \& Shaham (1985)
generalized these  works to Gaussian  random fields, showing  that for
scale-free power spectra ($P(k)  \propto  k^n$), dark matter halo  are
well  approximated by a    singular isothermal sphere ($\rho   \propto
r^{-2}$) for $-3<n<-1$, and that   the  density profile steepens   for
larger  value  of $n$.   N--body  experiments  confirmed roughly these
conclusions (Frenk  et  al.  1985; Quinn,  Salmon  \& Zurek 1986),  by
taking precisely into account   the complex dynamical behavior  of  3D
hierarchical clustering.

Observational estimates of gas and dark matter  mass profiles in X-ray
clusters have improved over  the past years  by  the emergence of  new
technics like strong and/or weak lensing analysis and temperature maps
obtained   by the X-ray emission.   Two  challenging properties of the
observed density profiles in X-ray clusters are  first that there is a
significant segregation  between gas and  dark matter (David, Jones \&
Forman 1995),  and second  that  mass estimates in  the inner  part of
clusters are not fitted by a singular isothermal sphere, but rather by
a much shallower profile, like a $r^{-1}$ law (Wu \& Hammer 1993) or a
very  small core  radius (Soucail   \& Mellier  1994). Very  recently,
Navarro, Frenk  \& White (1996) showed that  it was indeed the case in
high resolution N-body  simulations (see  also  Pfitzner 1996).   They
obtained a very general density profile well fitted by the formula

\begin{equation}
\rho(r) \propto \frac{1}{r(1+r/r_s)^2}
\label{navarro}
\end{equation}

\noindent with a single  shape parameter $r_s$.  

For the gaseous component, the situation is rather unclear.  Numerical
calculations performed by Pearce,  Thomas \& Couchman (1994)  revealed
that a small segregation between gas and dark matter occurs within the
hierarchical clustering scenario.  This would  be due to a  systematic
energy  transfer between  gas and  dark matter  during  mergers of the
small sub-clumps that  lead to  the final cluster.  On  the other hand,
Anninos \&  Norman (1996),   by  using a totally  different  numerical
method,  have  studied  gas and  dark  matter  density  profiles of  a
Coma-like  cluster  in a CDM   cosmogony.  They found  no  significant
segregation between gas and dark matter. Moreover, they found that the
dark matter density profile is not fitted by a Navarro  et al.  (1996)
model, but rather by a unique power law $\rho \propto r^{-9/4}$.

In this   paper,  we  address  these   questions  using an  analytical
approach.  This  has been possible  only  within a spherical  geometry
approximation, together with self-similar initial conditions. However,
we   are aware that spherical  collapse    is a strong  approximation,
because   it suffers  for  example  from  the so--called radial  orbit
instability (H\'enon  1973;  see also Binney  \&  Tremaine 1987).  Our
work mainly   extends the work  of  Fillmore \&   Goldreich (1984) and
Bertschinger (1985) to  the case  of  gas  collapse, and  discuss  the
consequences to more realistic  initial conditions.  We briefly recall
the initial conditions   and the basic notations  used  in Fillmore \&
Goldreich (1984)  to  derive  their  self-similar  solutions.    As we
restrict ourselves to  the spherical  case,  we simplify the  original
notations of Fillmore  \& Goldreich,  by taking their   dimensionality
$n=3$.

Each gas or dark matter  shell is labeled by  the mass $M_i$ initially
enclosed within its  initial radius $r_i$ at the  initial  time $t_i$. 
The initial velocity field is supposed to follow the Hubble law
 
$$
v_i = \frac{2}{3}\frac{r_i}{t_i}
$$

The  initial perturbation in mass  is assumed to be  a power law of the
initial mass

$$
\frac{\delta M_i}{M_i} = \left( \frac{M_i}{M_0} \right) ^{-\epsilon}
$$

Each spherical shell is expanding until the turn-around epoch, and the
trajectory   is described by  Kepler's  law  until it crosses  another
trajectory.   The  turn-around radius and  the turn-around  epoch of a
given shell $M_i$ are given by

\begin{equation}
r_* = \frac{r_i}{\delta} = r_0 
\left( \frac{M_i}{M_0} \right) ^{\epsilon+1/3}
\label{rta}
\end{equation}

\begin{equation}
t_* = \frac{3\pi}{4} \frac{t_i}{\delta ^{3/2}} = \frac{3\pi}{4} t_i
\left( \frac{M_i}{M_0} \right) ^{3\epsilon/2}
\label{tta}
\end{equation}

Another usefull quantity is the  current turn-around mass which labels
the current turn-around shell

$$
M(t) = \left( \frac{4}{3\pi} \right) ^{2/3\epsilon}
M_0 \left( \frac{t}{t_i} \right) ^{2/3\epsilon}
$$

Fillmore  \& Goldreich (1984)  derived self-similar  solutions for the
gravitational collapse  of  dark matter,  using these rather  peculiar
initial conditions.  Bertschinger (1985)  computed for  the particular
case $\epsilon=1$ the  solution for both gas  and dark matter. In this
paper, we derive  both gas and dark  matter self-similar solutions for
$0 < \epsilon \le 1$.  In order  to compare theoretically the dynamics
of each  component, we use a fluid  approximation to describe the dark
matter collapse.  We show in the first section that this approximation
is valid, because, as stated by Fillmore \& Goldreich (1984), the mass
$M(r,t)$   is an adiabatic  invariant.  Using  this fluid approach, we
derive  in  section  2 the  well-known  Fillmore  \& Goldreich  (1984)
self-similar  solutions in a totally   different way than the original
derivation. This gives us new insights in the nature of dark matter in
spherical geometry,  treated here   as a  one-dimensional  fluid.   We
derive also in  section 2 self-similar solutions  for the gas,  and we
find that they  differ from the dark  matter solutions.  In section 3,
we present numerical simulations to  validate our analytical treatment
of gas dynamics, and finally, we discuss  possible applications of our
work to other observational or numerical studies.

\section{A FLUID APPROACH FOR DARK MATTER}

In the original paper of Fillmore \&  Goldreich, the relaxed region of
the forming object, called  the halo, has  a mass profile parameterized
by

\begin{equation}
M(r,t) = \kappa (t) r^\gamma
\label{masse}
\end{equation}

\noindent where the function of time $\kappa$ was parameterized by 

\begin{equation}
\kappa  (t) \propto  t^{-s}.
\label{kappa}
\end{equation}

This is justified by the scale free nature of the problem, and this is
likely to be valid only in the most inner part  of the halo, where the
mass profile has reached asymptotically a power law both in radius and
in   time. Dark matter  particles in  the halo are  then following the
equation of motion

\begin{equation}
\frac{d^2r}{dt^2} = - \frac{GM}{r^2} = -G\kappa(t)r^{\gamma-2}
\label{eqom}
\end{equation}

Fillmore \&  Goldreich (1984) main hypothesis was   to assume that for
particles deep inside the halo, the function  $\kappa(t)$ was a slowly
varying function of time compared to one orbit period. This means that
the  right-hand  side of  equation  (\ref{eqom})   can be  taken as  a
function of   radius only over   one orbit  period,  and thus  we  can
integrate  equation  (\ref{eqom}) to obtain   an energy integral along
each trajectory

\begin{equation}
\left( \frac{dr}{dt} \right) ^2 = \frac{2G\kappa(t)}{\gamma-1}
\left(r_a^{\gamma-1} - r^{\gamma-1}\right)
\label{enint}
\end{equation}

\noindent where $r_a(t)$ is the apapsis radius of the current orbit. 
As we show it now, such an energy integral allows  to close the Vlasov
hierarchy which describes  the collision-less dynamics  of dark matter.

We define here  the  distribution function $f$ in  phase  space of the
dark matter particles, considering  only radial orbits. Therefore, the
distribution function   $f$ is  defined as  the  mass of  dark  matter
particles per unit  radius  $r$ and  per  unit  1D  velocity $v$,  and
satisfies the Vlasov equation

\begin{equation}
\frac{\partial f}{\partial t} + v\frac{\partial f}{\partial r} 
-\frac{GM}{r^2}\frac{\partial f}{\partial v} = 0
\label{vlasov}
\end{equation}

The fluid  limit of such a kinetic  description is obtained  by taking
the $n^{th}$ moments of equation (\ref{vlasov})  in velocity space. We
then obtain for the first three orders the mass conservation equation,
or continuity equation

\begin{equation}
\frac{\partial \mu}{\partial t} + 
\frac{\partial}{\partial r} \left( \mu u \right) = 0
\label{masscon}
\end{equation}

\noindent the momentum conservation equation, or Euler equation

\begin{equation}
\frac{\partial}{\partial t} \left( \mu u \right) + 
\frac{\partial}{\partial r} \left( \Pi+\mu u^2 \right) =
-\mu \frac{GM}{r^2}
\label{momcon}
\end{equation}

\noindent and the energy conservation equation

\begin{equation}
\frac{\partial}{\partial t} \left(\frac{1}{2}\Pi + \frac{1}{2}\mu u^2 
\right) + 
\frac{\partial}{\partial r} \left(\frac{3}{2}\Pi u+\frac{1}{2}\mu u^3 
\right) =
-\mu u \frac{GM}{r^2} + \frac{\partial S_3}{\partial r}
\label{enercon}
\end{equation}

\noindent The mass per unit length is noted $\mu$, the mean velocity  
of dark matter particles in a given fluid  element $<v>$ is noted $u$,
and  $\Pi$  is related  to the   velocity  dispersion  of  dark matter
particles  in  this fluid   element  by $\Pi   =  \mu  (<v^2>-u^2)  $.
Therefore, $\Pi$  is the analogous of  a thermal pressure in the dark
matter fluid.  This hierarchy is a  priori not closed,  and one has to
know the exact form of the third moment $S_3$ to solve the problem.

Using the energy integral (eq.    [\ref{enint}]), it appears that  the
fluid element  at position $r$  is crossed by particles  with opposite
velocities which  follow trajectories labeled  by their apapsis.  The
distribution function is therefore even in velocity  space, and  moments
of odd order vanish.  This implies that $S_3
\simeq 0$ and also $u \simeq 0$.

The  fluid approximation ($S_3  = 0$)  allows us  to  close the set of
equations (\ref{masscon}-\ref{enercon})    which becomes    the  fluid
equations for  dark matter.   Since   these equations are  hyperbolic,
discontinuities in the flow appear  naturally.  They are treated using
the   Rankine-Hugoniot   discontinuity  relations.  Physically,   bulk
kinetic  energy $1/2\mu u^2$ is  thus dissipated into internal kinetic
energy  $1/2\mu  <v^2>$.  Relaxation   in  the  dark matter fluid   is
equivalent to shell crossings  and  the discontinuity is analogous  to
the  first caustic in  the dark matter density  profile.  In the exact
treatment   of Fillmore \& Goldreich   (1984),  relaxation occurs in a
thick  layer,    between the first     caustic   (where $S_3$  differs
significantly from zero) and the radius  where the fluid approximation
is roughly recovered ($S_3 \simeq 0$).

\section{GAS AND DARK MATTER SIMILARITY SOLUTIONS}

Our fluid approach gives new insight in dark matter dynamics.  It also
allows us to compare  directly gas and dark  matter dynamics.  Because
in our fluid approximation we  don't track particles trajectories,  it
is possible to  derive the self-similar  solutions in a  rather simple
analytical way.   In fact, the  mass enclosed  by  a given  Lagrangian
fluid element    remains  constant in time.    Each  fluid  element is
initially labeled by the initial  enclosed  mass $M_i$.  Because  this
mass  remains constant, we  drop here the   subscript $i$ and we label
each fluid element by its enclosed mass $M$.

We first consider   the case of  the pure  dark  matter  collapse, and
recover Fillmore \& Goldreich  results, and finally we investigate the
pure  gas collapse,  showing that the  halo  profile in both cases can
differ significantly.

\subsection{Dark Matter Halo Profiles in the Fluid Approximation}

Each fluid element first  expands, and reaches turn-around at $t=t_*$.
After  turn-around,  it  falls towards   the  center and  reaches  the
relaxation front. Self-similarity  implies that this  front is located
at a constant fraction of the turn-around radius

$$
r_s = \lambda_s r_*
$$

\noindent and that the fluid element get shocked at a constant fraction
of the turn-around epoch 

$$
t_s = \tau_s t_*
$$

Behind  the shock, the fluid element  gradually reaches the asymptotic
regime.  During this self-similar regime, we parameterize its evolution
by

\begin{equation}
r(M,t) \propto M^{p} t^{-q}
\end{equation}

\noindent where $q \ge 0$. To ensure self-consistency, we extrapolate 
this regime back   to  $t = t_s$.    We  then obtain   using equations
(\ref{rta}) and (\ref{tta})

\begin{equation}
p = \frac{1}{3} + \epsilon + \frac{3}{2}q\epsilon
\label{equonp}
\end{equation}

Mass conservation implies that $\gamma =  1/p$ (eq. [\ref{masse}]) and
$s = -q/p$   (eq.  [\ref{kappa}]).   We  then assume  that each  fluid
element is in hydrostatic   equilibrium.  In the previous  section, we
noticed that  this  assumption  was  valid  during  the asymptotic halo
regime. Note however that it doesn't imply  necessarily that $q=0$. We
will see  in the followings, that  it is in fact  possible to obtain a
quasi-hydrostatic  flow,      which  is  not     strictly stationary.

Hydrostatic equilibrium writes then for a fluid element $M$

\begin{equation}
\frac{\partial \Pi}{\partial M} = -\frac{GM}{r^2}
\label{hydrostaticdark}
\end{equation}

\noindent Integrating this equation between $M$ and the currently 
shocked fluid element  $M_s(t)$, and  using the  scaled variable  $x =
M/M_s(t)$, leads after a few algebra to

\begin{equation}
\Pi (M,t) = \Pi_s(t) + \frac{GM_s(t)^2}{R_s(t)^2} 
\int _{M/M_s}^1 x^{1-2p}dx 
\label{virieldark}
\end{equation}

\noindent which is consistent with  the Virial theorem.  The  integral
in equation (\ref{virieldark}) converges or  diverges whether $p <  1$
or $p \ge 1$.

Let us assume first that  $p < 1$.  In  the asymptotic regime, one has
$M/M_s(t) \ll 1$. The integral in  equation (\ref{virieldark}) is thus
a constant to   leading order.  Consequently,   in the  halo $\Pi$  is
uniform, and is determined entirely by the post-shock conditions.  The
hydrostatic equilibrium equation  cannot be  satisfied  in that  case.
The validity range for $p$ is therefore $p \ge 1$.  We deduce then the
validity range for $q$ from equation (\ref{equonp}).

$$
\epsilon \ge \frac{2}{3} ~~~~~ q \ge 0 
$$

\begin{equation}
\epsilon \le \frac{2}{3} ~~~~~ q \ge \frac{2}{3\epsilon}\left(\frac{2}{3}
-\epsilon\right) 
\label{validonq}
\end{equation}

As $p \ge 1$, the integral diverges, and to leading order, one gets

$$
\Pi (M,t) = \frac{GM_s(t)^2}{R_s(t)^2} \left( \frac{M}{M_s(t)} \right) 
^{2-2p}
$$

It is straightforward to   check that the hydrostatic   condition (eq.
[\ref{hydrostaticdark}]) is fulfilled  in that case.  We need another
equation to find the final solution  of the problem. The last equation
on $\Pi$  states that twice the  internal energy of  a  given shell is
equal to its gravitational energy.  The action is thus given by

$$
S (M,t) = \int _{t_*} ^t \left( \frac{dr}{dt} \right) ^2 dt
$$

Integrating  by part  and    using again the hydrostatic   equilibrium
assumption leads to

$$
S (M,t) =  q \frac{r_*^2}{t_*} 
\left\{  1 - \left(\frac{t}{t_*}\right)^{-2q-1} \right\}
$$

\noindent Since  $q \ge 0$,  the  action is always  a constant in  
the asymptotic regime where $t \gg t_*$, and takes the value

\begin{equation}
S (M) = q \frac{r_*^2}{t_*}
\end{equation}

\noindent Applying the  least action principle  means here that  
the solution  is obtained for the minimum  value of $q$. From equation
(\ref{validonq}), we find

$$
\epsilon \ge \frac{2}{3} ~~~~~ q = 0
$$

$$
\epsilon \le \frac{2}{3} ~~~~~ q = \frac{2}{3\epsilon}
\left(\frac{2}{3}-\epsilon\right)
$$

\noindent In this way, we   finally  recover  the   solution  of    
Fillmore  \& Goldreich  within  our  fluid  approximation.  The   halo
profile is therefore given by

$$
\epsilon \ge \frac{2}{3} ~~~~~ \gamma = \frac{3}{3\epsilon+1} ~~~~ s=0 
$$

$$
\epsilon \le \frac{2}{3} ~~~~~ \gamma = 1 ~~~~ s=\frac{2}{3\epsilon}
\left(\epsilon-\frac{2}{3}\right)
$$

For  $\epsilon    \ge   2/3$,   the    density   profile  scales    as
$r^{-9\epsilon/(3\epsilon+1)}$,        and  is  strictly    stationary
($s=0$). For  $\epsilon \le 2/3$, the density  scales as $r^{-2}$, and
increases continuously with time.  Moreover,  we deduced also from the
hydrostatic equilibrium equation  that $r^{-2}$ is the flattest stable
density profile which can be obtained with purely radial orbits.

\subsection{Gas Halo Profiles}

We   now  turn to the   pure gas  collapse.   The   derivation that we
presented in the  previous section for  dark matter can be  applied to
the  gas  dynamics, though the  gas has  an  isotropic distribution in
velocity space.

Hydrostatic equilibrium writes for a gas fluid element $M$

\begin{equation}
4\pi r^2\frac{\partial P}{\partial M} = -\frac{GM}{r^2}
\label{hydrostaticgas}
\end{equation}

\noindent where $P$ is now the pressure  of the  gas.   Integrating  
again   this equation between $M$  and    the currently shocked  fluid
element $M_s(t)$ leads to

\begin{equation}
P (M,t) = P_s(t) + \frac{GM_s(t)^2}{4\pi R_s(t)^4} 
\int _{M/M_s}^1 x^{1-4p}dx 
\label{virielgas}
\end{equation}

\noindent where we use the same scaled variable as before $x = M/M_s(t)$.
Note that this   last  equation differs  from  the corresponding  dark
matter   equation (eq.   [\ref{virieldark}]),  due  to   the different
dimensionality of the phase space.  The integral converges or diverges
now if $p < 1/2$ or $p \ge 1/2$. The case $p < 1/2$ has to be rejected
in order  to satisfy the  hydrostatic equilibrium assumption.  We then
find the validity range for $q$

$$
\epsilon \ge \frac{1}{6} ~~~~~ q \ge 0 
$$

$$
\epsilon \le \frac{1}{6} ~~~~~ q \ge \frac{2}{3\epsilon}\left(\frac{1}{6}
-\epsilon\right) 
$$

\noindent The action takes the same form as for dark matter, and 
again, the least action  principle applied on  each gas  fluid element
leads to the self-similar solutions for gas

$$
\epsilon \ge \frac{1}{6} ~~~~~ \gamma = \frac{3}{3\epsilon+1} ~~~~ s=0 
$$

$$
\epsilon \le \frac{1}{6} ~~~~~ \gamma = 2 ~~~~ s=\frac{4}{3\epsilon}
\left(\epsilon-\frac{1}{6}\right)
$$

For  $\epsilon    \ge     1/6$,  the  density    profile    scales  as
$r^{-9\epsilon/(3\epsilon+1)}$,  and is   strictly stationary ($s=0$).
For $\epsilon \le 1/6$, the density scales  as $r^{-1}$, and increases
continuously with  time.  These  new solutions  differ  from the  dark
matter ones, although they show the same two characteristic regimes: a
strictly stationary and hydrostatic flow for $\epsilon \ge 1/6$, and a
non-stationary, but quasi-hydrostatic flow for $\epsilon \le 1/6$.  We
also show that the flattest stable density profile is $r^{-1}$ in that
case.

\section{NUMERICAL SOLUTIONS}

This analytical approach can be  tested numerically.  For dark matter,
Fillmore \& Goldreich (1984) solved  the Vlasov-Poisson equations  and
described     in  a semi-analytical   way   the  dark matter particles
trajectories.  They confirmed  their  analytical work,  and therefore,
the validity of the dark  matter self-similar solutions.  Bertschinger
(1985)  solved the pure gas collapse  for $\epsilon  = 1$, integrating
semi-analytically the gas dynamics  equations.  He found that  the gas
density profile was similar  to the corresponding dark matter profile.
We recovered also this result in the previous section.

To  test our  gas  self-similar solutions, we  use here  the spherical
hydrodynamical code, presented  in Chi\`eze, Teyssier \& Alimi (1996),
to which  the reader is referred.  Shock  waves are followed using the
pseudo--viscosity method, with a  tensorial formulation of the viscous
stress (Tscharnuter \& Winkler 1979; Chi\`eze et al.  1996). The shock
front   is captured within two   or   three cells, without  post-shock
oscillations.   These calculations allow  us to recover the asymptotic
halo  regime, and also  the exact  location  of the  shock front.  The
self--similar  initial   conditions  are introduced   in   our code as
follows: the general power--law  density profile is smoothly connected
to a  non--singular,  inner core,  containing  less than 0.1\% of  the
final turn--around mass.  We then choose an initial epoch, such as the
density  contrast in  the  center is  equal to  1\%.  The initial  gas
temperature  is uniform and  equal to the  cosmic background radiation
temperature.  This is in  any case a very  small fraction of the final
halo temperature  (typically $10^{-8}$).  The hydrodynamical equations
are then solved using physical coordinates ($r$, $t$), rather than the
self--similar coordinate $\lambda(t)$.

\begin{figure}[httb]
\hbox{\psfig{file=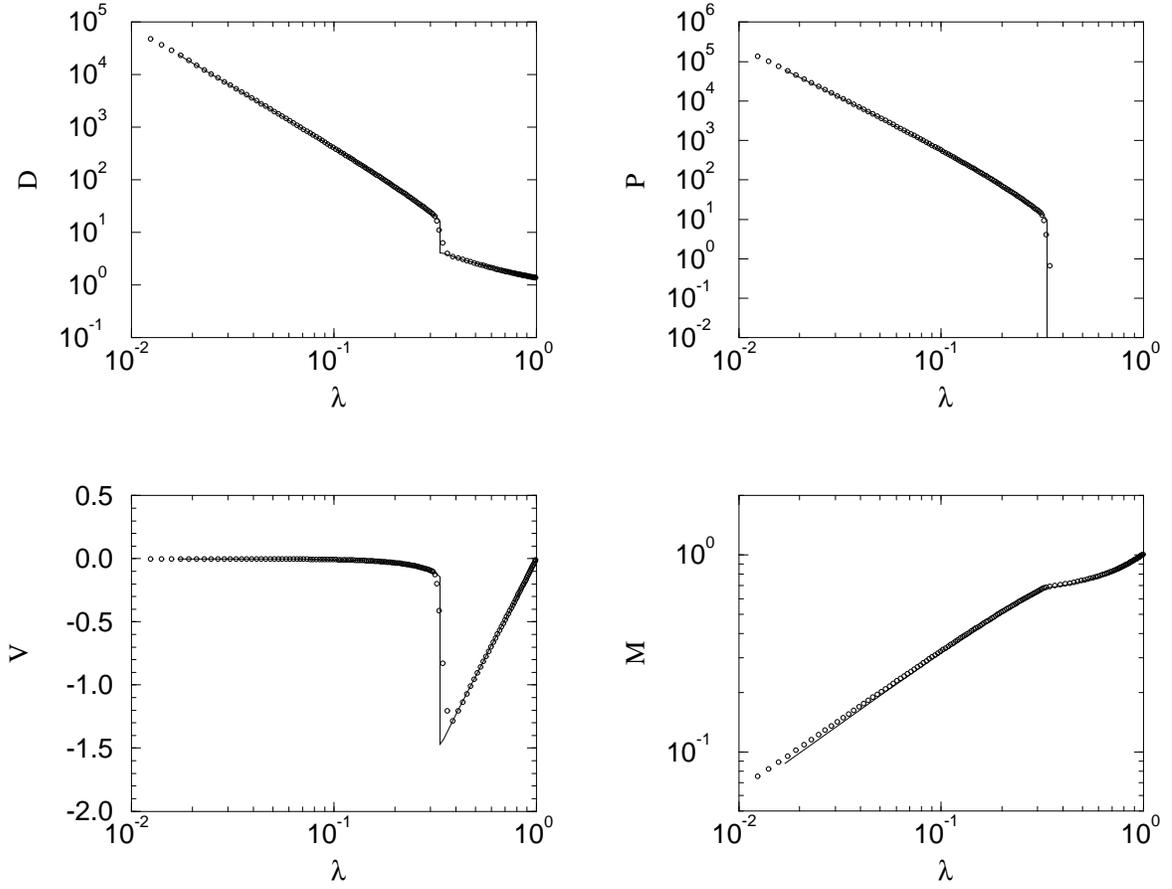,height=13cm,angle=270}}
\caption{Density, pressure, velocity and mass profiles obtained in the
$\epsilon=1$ case for   the pure gas   collapse, as a  fonction of the
self--similar radius $\lambda = r/r_* (t)$ (see text). Solid lines are
the results obtained  by Bertschinger  (1985) using a  semi-analytical
method.  Open  circle are the  results obtained  by our hydrodynamical
code.  Both results agree perfectly well. }
\label{bertschinger}
\end{figure}

\begin{figure}[httb]
\hbox{\psfig{file=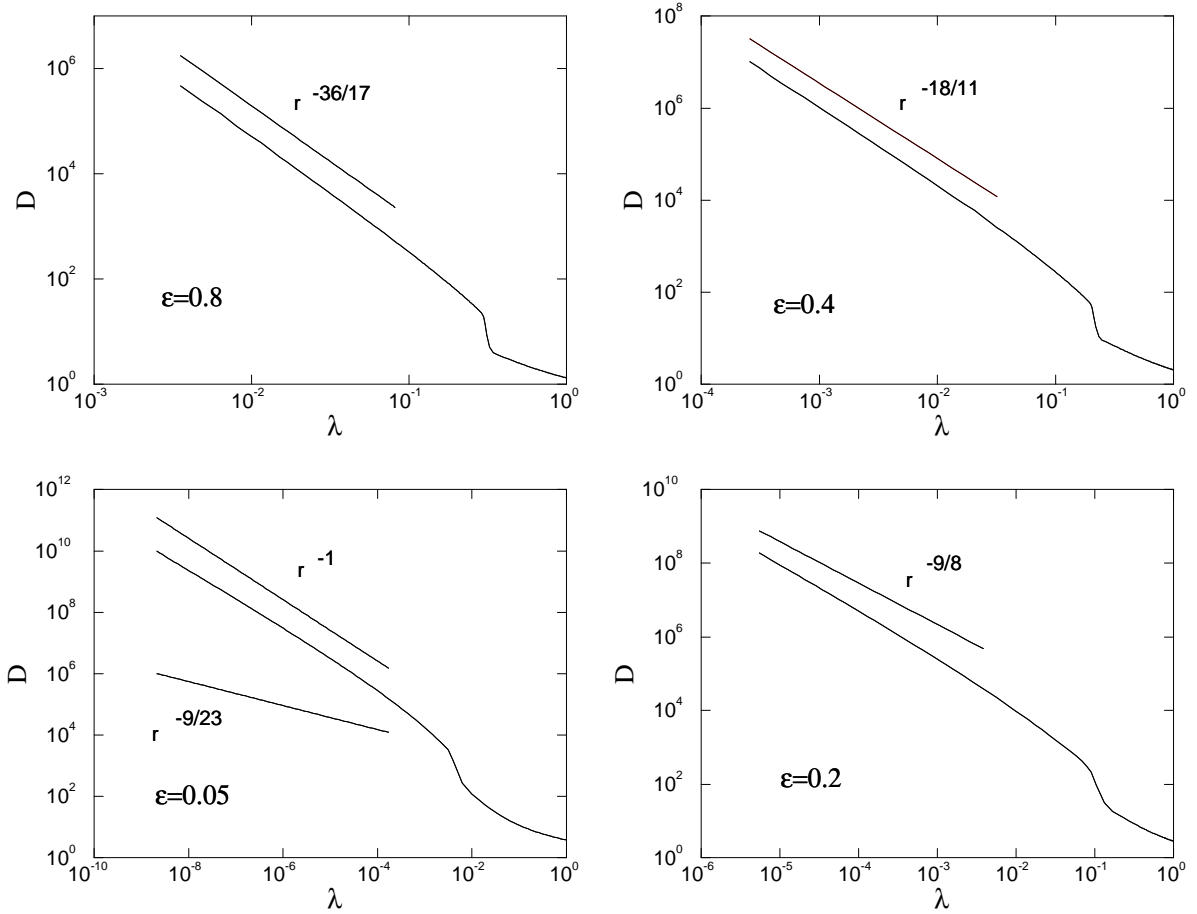,height=13cm,angle=270}}
\caption{Density profiles obtained  for  $\epsilon=$0.8, 0.6,  0.4 and
0.05 by our  hydrodynamical code, as  a fonction  of the self--similar
radius $\lambda = r/r_* (t)$. The  straight line in  each graph is the
power  law predicted by the analytical  calculations performed in this
paper. }
\label{profiles}
\end{figure}

In figure  (\ref{bertschinger}), we plot  the  mass, density, velocity
and  pressure profiles obtained for  the $\epsilon=1$ run.  We compare
our results with the semi-analytical profiles obtained by Bertschinger
(1985).  Note that both results agree  remarkably well.  Moreover, the
computed  shock front position is recovered  within  3\% accuracy.  In
figure (\ref{profiles}), we plot  the  density profiles  obtained  for
$\epsilon$ = 0.8, 0.6, 0.4 and 0.05.  The  straight line shows in each
graph the corresponding  asymptotic power law derived  analytically in
the last section.  The numerical results agree perfectly well with the
analytical ones.  The $\epsilon=0.05$  case illustrates the change  of
behavior in the  halo  dynamics ($\epsilon \le  1/6$).   The power law
$\rho  \propto   1/r$ is   very well   recovered.  We list    in table
(\ref{tableeps}) the shock  front position $\lambda_s$ computed by our
hydrodynamical code.  Note that  for lower $\epsilon$, the shock front
is located at a much deeper radius.

\section{DISCUSSION}

We  now  discuss possible    applications of  our  results on   cosmic
structures formation theory.  We discuss here three major consequences
of our work : an origin for segregation between gas and dark matter in
X-ray clusters (David   {\it et al.}   1996),  an explanation of   the
Navarro et  al.  (1996) density  profile and  a justification for  the
stable clustering hypothesis.

\subsection{Mass Distribution in the Core of X-Ray Clusters}

The  first application    of the present   work   is  to propose    an
interpretation  of Navarro {\it et   al.}  (1996) dark matter  density
profile found  with  high  resolution   N-body simulations (see    eq.
[\ref{navarro}]).  This profile scales   as $r^{-1}$ in the center  of
the cluster, as  $r^{-2}$ for a  large range of intermediate radii and
as $r^{-3}$ in the  outer regions.  The  outer regime, located roughly
at the Virial radius, is likely to be the relaxation layer where shell
crossings dominate, which leads to  a profile steeper than isothermal.
The intermediate power law ($r^{-2}$)  is expected from the analytical
approach  for  $\epsilon \le  2/3$  and   for purely  radial   orbits.
Moreover, Tormen,   Bouchet   \& White (1996)  recently    studied the
distribution  function in  velocity   space of simulated  dark  matter
halos.  They calculate for  that purpose the factor $\beta  (r) =  1 -
\sigma^2_t/2\sigma^2_r$  where     $\sigma_t$  and   $\sigma_r$    are
respectively the tangential  and the radial  velocity  dispersion at a
given radius.  $\beta = 1$ corresponds to purely radial orbits; $\beta
=  0$ to  isotropic orbits and  $\beta  = -\infty$  to purely circular
orbits.  In  the  region where the   density scales as $r^{-2}$,  they
found $\beta \simeq 1$, while in the central region, where the density
scales as $r^{-1}$,  they found $\beta  \simeq 0$.  In  this paper, we
show that  for $\beta  = 1$, the   flattest stable density  profile is
$r^{-2}$, and for the gas, which has $\beta  = 0$, the flattest stable
density  profile is $r^{-1}$.   The  results  presented in this  paper
suggest that the change of slope in the density  profile may be due to
a change of dimensionality in the dark matter velocity space.

\subsection{Segregation between Gas and Dark Matter}
 
We showed that  for $\epsilon \le  2/3$, the density profiles obtained
in the pure dark matter  case differ from those  obtained in the  pure
gas collapse.  In cosmic  structures, we expect  both fluids  (gas and
dark   matter) to  be dynamically important.    Therefore, the coupled
dynamics has to be  studied in order to have  definitive answers.   We
have not been able to solve here analytically the coupled case, mainly
because  self-similarity  is broken by  the presence   of two distinct
fluids.  The  total mass profile is  however expected to  have a power
law, between  the  pure gas and  the  pure dark  matter profiles  with
$3/(1+3\epsilon) <  \gamma < 1$.  The exact  value of $\gamma$ depends
on the  value of $\Omega_B$, the  universal gas fraction.  In Chi\`eze
{\it et al.} (1996), we study the case of the single spherical Fourier
mode.  It can be shown that this corresponds  to $\epsilon = 4/9$.  As
expected, we obtain a segregation which depends on the chosen value of
$\Omega_B$.  Bertschinger (1985) studied  the case $\epsilon=1$ in the
coupled case, assuming  a very low  value for $\Omega_B$. As expected,
he  found   no segregation.  Our  belief   is that there  is  always a
segregation between gas and dark matter for $\epsilon \le 2/3$, and no
segregation for  $\epsilon \ge  2/3$.  This  would be due  to the fact
that the dark   matter evolution is   not strictly stationnary  in the
halo.  This  results in a systematic delay  of gas shells  relative to
dark  matter  shells.   Following  Hoffman  \&  Shaham (1985),  it  is
possible  to find the  most probable initial  density profile around a
peak in the initial Gaussian random field.   This leads to a power law
initial  perturbation which  is   roughly proportional  to the  linear
two--point correlation function with $\epsilon = (n+3)/3$.  We propose
here  that power spectra  with $n \le -1$  lead to segregation between
gas and dark matter, with a dark matter  density profile $\rho \propto
r^{-2}$ and a gas density  profile slightly shallower.  However, if $n
\ge -1$, gas and  dark  matter are  expected to have  similar  density
profiles  with $\rho \propto r ^{-9  \epsilon / (1+3\epsilon)}$.  This
rather  strong  prediction has   to be  tested   in  3D hydrodynamical
simulations.   Anninos  \&   Norman    (1996) found   no   significant
segregation  in their 3D simulations.   They used a standard CDM power
spectrum where $-1 < n < 0$  on clusters scales.  Moreover, they found
that the halo  density profiles are well  fitted by a $r^{-9/4}$ power
law, both for gas and dark matter.

\subsection{Stable Clustering Hypothesis}

The last consequence  of our work  concerns one of  the most powerful
analytical   prediction in gravitational   dynamics, namely the stable
clustering   ansatz (Peebles 1980).   This hypothesis  states that on
small scales the relative pair   velocity cancels exactly the  Hubble
flow.  For   scale-free power spectra, this  leads   to the non-linear
two-point correlation function $\xi \propto r^{-\gamma}$ with $\gamma
= (9+3n)/(5+n)$.  This regime  is valid for  $\xi \gg 1$ and has been
extensively tested by N-body  simulations (Jain 1996).  Padmanabhan et
al.   (1995) try to find another  derivation  of the stable clustering
hypothesis using the  secondary infall model (which  is the main topic
of our paper).  Instead of the  linear two-point correlation function,
they  used  for  the  initial   spherical  perturbation the   smoothed
r.m.s. density contrast. This leads to $\epsilon = (n+3)/6$. Using the
solution of Fillmore \& Goldreich  (1984), the resulting halo profiles
are (Jain 1996)

$$
n \ge 1 ~~~~ \rho \propto r^{-(9+3n)/(5+n)}
$$
$$
n < 1  ~~~~  \rho \propto r^{-2}
$$

The next step is to assume that the  two-point correlation function in
the highly non-linear regime is dominated by the density field of such
spherical halos.  This  has led  Jain   (1996) to conclude  that  the
stable clustering hypothesis   might  be invalid for   $n<1$. However,
Fillmore \&  Goldreich  (1984) results  apply only  for  purely radial
orbits.  Therefore, the  stable  clustering regime might  be recovered
within a fully 3D  density field. We  saw in  section 5.1 that  recent
N-body experiments suggest that on  very small scales the distribution
of dark  matter  particles  is  nearly isotropic  in   velocity space.
Therefore, to derive the halo profiles using the initial conditions of
Padmanabhan et al.  (1995), one has to use  the new solutions we found
in this paper.  This reads

$$
n \ge -2 ~~~~ \rho \propto r^{-(9+3n)/(5+n)}
$$
$$
n < -2  ~~~~  \rho \propto r^{-1}
$$

Therefore,   assuming that  the    two-point correlation  function  is
dominated in the highly non-linear  regime by the density profiles  of
quasi-spherical halos with an isotropic  distribution of particles in
velocity space, one could justify the stable clustering hypothesis for
$n \ge -2$.  Significant deviation could  be detected for $n = -2.5$,
but this  rather flat power spectra might  be difficult to study using
N-body simulations.

\subsection{Conclusions}

In this paper, we derive analytically gas and dark matter self-similar
solutions for  the gravitational collapse  of a spherical perturbation
embedded in an expanding Universe.  The gas self-similar solutions can
be   easily extended  to   dark matter  particles   with an  isotropic
distribution in velocity space.  The behavior of the gas differs from
the dark matter one, leading to different  results for each fluid.  We
find three possible consequences of our work  : a possible segregation
between gas and dark matter for a  certain class of initial conditions
($\epsilon \le 2/3$), a possible explanation for a dark matter density
profile $\rho  \propto  r^{-1}$    in the central   region  of  cosmic
structures,  and a  possible   justification of the  stable clustering
hypothesis for power spectra with $n \ge -2$.

\pagebreak

\begin{table}[httb]
\begin{center}
\begin{tabular}{| c || c | c | c | c | c | c |}
\hline
$\epsilon$         & 1    & 0.8  & 0.6  & 0.4  & 0.2  & 0.05 \\
\hline 
$\lambda_s$        & 0.36 & 0.33 & 0.29 & 0.23 & 0.13 & 6.2$\times 10^{-3}$\\
\hline
$\Delta \lambda_s$ & 0.01 & 0.01 & 0.01 & 0.01 & 0.02 & 3.0$\times 10^{-3}$\\
\hline
\end{tabular}
\end{center}
\caption{Self similar radius of the shock front and the corresponding 
uncertainties due to finite resolution obtained for 
different values of $\epsilon$ by our hydrodynamical code.}
\label{tableeps}
\end{table}

\pagebreak

\pagebreak

\begin{thebibliography}{}
  
\bibitem[Anninos  \& Norman  1996]{Anninos96}Anninos, P.A., \& Norman,
  M.L., 1996, ApJ, 459, 12
 
\bibitem[Bertschinger    1985]{Bertschinger85} Bertschinger, E., 1985,
  ApJS, 58, 39
  
\bibitem[Binney \& Tremaine   1987]{BT87}Binney  J., \&  Tremaine  S.,
 1984, Galactic Dynamics, Princeton University Press


\bibitem[Chi\`eze,  Teyssier and  Alimi 1996]{CTA96} Chi\`eze, J.-P.,
  Teyssier, R., \& Alimi, J.-M., 1996, submitted to ApJ.
  
\bibitem[David, Jones \& Forman 1995]{David95} David, L.P., Jones, C.,
\& Forman, W., 1995, ApJ, 445, 578.

\bibitem[Fillmore  \&     Goldreich    1984]{FG84}Fillmore, J.A.,   \&
  Goldreich, P., 1984, ApJ, 281, 1 (FG84)

\bibitem[Frenk et al. 1985]{Frenk85} Frenk, C.S., White, S.D.M., 
  Efstathiou, G.P., \& Davis, M. 1985, Nature, 317,595

\bibitem[Gunn \& Gott 1972]{GG72} Gunn, J. \& Gott, J.R. 1972, 
  ApJ, 209, 1

\bibitem [H\'enon 1973] {H73} H\'enon, M., 1973, A\&A, 24, 229.

\bibitem[Hoffman \& Shaham 1985]{HS85} Hoffman, Y. \& Shaham, J. 1985, 
  ApJ, 297, 16

\bibitem[Jain 1996]{J96}Jain, B., 1996, astro-ph/9605192

\bibitem[Navarro,  Frenk  \&  White  1996]{Navarro96} Navarro,   J.F.,
  Frenk, C.S., White, S.D.M., 1996, Apj, 462, 563

\bibitem[Padmanabhan et al. 1995]{P95}Padmanabhan, T., Cen, R., Ostriker, 
  J.P., Summers, F.J., 1995, astro-ph/9510037

\bibitem[Pearce, Thomas, \& Couchman 1994]{Pearce94} Pearce, F., Thomas, 
  P.A.  \& Couchman, H.M.P., 1994, MNRAS, 268,953.

\bibitem[Peebles 1980]{P80} Peebles, P.J.E., 1980, The Large-Scale Structure of
the Universe (Princeton: Princeton University Press)

\bibitem [Pfitzner  1996]   {P96} Pfitzner, D.,   1996,  procedings of
  ``Second  Stromlo    Symposium  on the    Structure  and Dynamics of
  Elliptical  Galaxies'', Mont Stromlo  Observatory,  August  1996, in
  press.

\bibitem[Quinn, Salmon \& Zurek 1986]{QSZ86} Quinn, P.J., Salmon, J.K. 
  \& Zurek, W.H. 1986, Nature, 322, 329

\bibitem[Soucail \& Mellier (1994)]{SM94}Soucail, G., \& Mellier, Y. 
  1994, in Gravitational  Lenses  in the Universe,  ed. J.Surdej  et al.
  (li\`ege: Univ. of Li\`ege), 595

\bibitem[Tormen, Bouchet  \& White  1996]{Tor96} Tormen,  G., Bouchet,
  F.R., \& White, S.D.M., 1996, astro-ph/9603132, submitted to MNRAS.

\bibitem[Tscharnuter \& Winkler (1979)]{Tschar79}Tscharnuter, W.-M.,  \&
  Winkler, K.-H., 1979, Comput. Phys. Comm., 18, 171

\bibitem[Wu \& Hammer 1993]{Wu93} Wu, X.P., \& Hammer, F. 1993, MNRAS,
  262, 187.
  
\end{thebibliography}
\end{document}